\documentclass[aps,preprint]{revtex4}%
\usepackage{amsfonts}
\usepackage{amsmath}
\usepackage{amssymb}
\usepackage{graphicx}%
\begin{document}
\title{{\LARGE COSMOLOGICAL MODEL FOR THE VERY EARLY UNIVERSE IN B.D. THEORY}}
\author{Marcelo Samuel Berman }
\affiliation{Editora Albert Einstein \ Ltda\ \ \ }
\affiliation{Av. Candido Hartmann, 575 - \# 17 \ \ }
\affiliation{80730-440 - Curitiba - PR - Brazil}
\begin{abstract}
Following a paper by Berman and Marinho Jr (2001), where it was established an
equation of state ($p=-\frac{1}{3}\rho$), for the very early Universe, under
which, Einstein's equations with $\Lambda=0$\ , render a scale-factor
proportional to the \ time coordinate, and under which, Berman(2006, 2006a)
showed that in the $\Lambda\neq0$\ \ case, the Universe entered automatically
into an exponential inflationary phase, we now study the corresponding effect
in Brans-Dicke Cosmology. We find that the resulting models are similar to
those in General Relativity Cosmology.\ \ \ \ \ 

\end{abstract}
\maketitle

\begin{center}
{\LARGE COSMOLOGICAL MODEL FOR THE VERY EARLY UNIVERSE IN B.D. THEORY}

\bigskip

Marcelo Samuel Berman
\end{center}

\bigskip

\bigskip Berman and Marinho Jr(2001), proved that, Brans-Dicke relation (Brans
and Dicke, 1961), in the form, \ 

$\frac{GM}{R}=\gamma=$\ \ constant \ \ \ ,\ \ \ \ \ \ \ \ \ \ \ \ \ \ \ \ \ \ \ \ \ \ \ \ \ \ \ \ \ \ \ \ \ \ \ \ \ \ \ \ (1)\ \ \ 

\bigskip

where $\gamma$\ \ is of order unity, results in the equation of state,

\bigskip

$p=-$ $\frac{1}{3}\rho$\ \ \ \ \ \ \ . \ \ \ \ \ \ \ \ \ \ \ \ \ \ \ \ \ \ \ \ \ \ \ \ \ \ \ \ \ \ \ \ \ \ \ \ \ \ \ \ \ \ \ \ \ \ \ \ \ \ \ (2)\ \ \ 

\bigskip

In the above two relations, \ $M$\ , \ $R$\ , \ $p$\ \ and \ $\rho$\ \ , stand
respectively for the mass and radius of the causally related Universe ($R$
\ is identified altogether with the scale-factor in Robertson-Walker's
metric), cosmic pressure, and energy density. In consequence, the above
equation of state, valid immediately after Planck's time does not lead to a
radiation-dominated Universe. That relation \ (1) is valid for the present
Universe, there is no doubt (Weinberg, 1972); that it is valid for the
Planck's Universe, there is also no doubt (Berman, 1994). The obvious
generalization \bigskip(Berman and Marinho Jr, 2001),\ is to suppose that in
the very early Universe, following Planck's time, the relation continued to be
valid, because it should not be just a coincidental relation, valid just for
that particular instant of time.

\bigskip

With\ equation of state (2), that was derived for the very early Universe,
Berman and Marinho Jr\ (2001) obtained a scale-factor proportional to the
time\ \ coordinate, in the

$\Lambda=0$\ \ case, where \ $\Lambda$\ \ stands for the (zero-valued)
cosmological constant.\ When \ $\Lambda\neq0$\ , Berman(2006, 2006a)\ \ has
shown that an exponential inflationary phase (Guth, 1981), does arise, as a
consequence of the above equation of state. Of course, we are dealing with
Robertson-Walker's metric:

\bigskip

$ds^{2}=dt^{2}-\frac{R^{2}}{\left[  1+\frac{k\text{ }r^{2}}{4}\right]  ^{2}%
}d\sigma^{2}$\ \ \ \ \ \ .\ \ \ \ \ \ \ \ \ \ \ \ \ \ \ \ \ \ \ \ \ \ \ \ \ \ \ \ \ (3)\ \ 

\bigskip

In the causally related Universe, with the volume, \ $V=\alpha R^{3}$\ , we
have, with \ $\alpha=$\ \ constant, \ 

\bigskip

$M=\rho V=\alpha R^{3}\rho$ \ \ \ \ \ \ \ \ \ \ .\ \ \ \ \ \ \ \ \ \ \ \ \ \ \ \ \ \ \ \ \ \ \ \ \ \ \ \ \ \ \ \ (4)

\bigskip

In this case (Berman, 2006, 2006a),

\bigskip

\bigskip$\gamma=G\alpha\rho R^{2}$ \ \ \ \ \ \ \ \ \ \ \ \ \ \ \ \ \ . \ \ \ \ \ \ \ \ \ \ \ \ \ \ \ \ \ \ \ \ \ \ \ \ \ \ \ \ \ \ \ \ \ \ \ \ \ (5)

\bigskip

We now turn our attention to the Brans-Dicke field equations derived by Uehara
and Kim (1982) with a cosmological constant, which read:

\bigskip

$\frac{\ddot{\phi}}{\phi}+3H\frac{\dot{\phi}}{\phi}-\frac{4\Lambda}{3+2\omega
}=\frac{\kappa}{\left(  3+2\omega\right)  \phi}(\rho-3p)$\ \ ,\ \ \ \ \ \ \ \ \ \ \ \ \ \ \ \ \ \ \ \ \ \ \ \ (6)

\bigskip

$\dot{\rho}=-3H(\rho+p)$ \ \ \ \ \ \ \ \ \ \ \ \ \ \ \ \ \ \ \ \ \ \ \ \ \ \ \ \ \ \ ,\ \ \ \ \ \ \ \ \ \ \ \ \ \ \ \ \ \ \ (7)

\bigskip

$H^{2}+kR^{-2}=\frac{\kappa\rho}{3\phi}-H\frac{\dot{\phi}}{\phi}+\frac{\omega
}{6}\left(  \frac{\dot{\phi}}{\phi}\right)  ^{2}+\frac{\Lambda}{3}$ \ ,\ \ \ \ \ \ \ \ \ \ \ \ \ \ \ \ \ \ \ \ \ (8)

\bigskip where \ $\kappa=8\pi$\ \ .

\bigskip

It is important to remember, that from Solar tests, the following relation\ is valid:

\bigskip

$G=\phi$ $\left[  \frac{2\omega+4}{2\omega+3}\right]  $\ \ \ \ \ \ \ \ \ \ \ \ \ \ \ \ \ \ .\ \ \ \ \ \ \ \ \ \ \ \ \ \ \ \ \ \ \ \ \ \ \ \ \ \ \ \ \ \ \ \ \ \ \ \ (9)

\bigskip

Taking care of relation (5), one finds the following solution:

\bigskip

$H=\frac{\dot{R}}{R}=$ \ constant \ \ \ \ \ \ . \ \ \ \ \ \ \ \ \ \ \ \ \ \ \ \ \ \ \ \ \ \ \ \ \ \ \ \ \ \ \ \ \ \ \ \ \ (10)

\bigskip

$R(t)=R_{0}$ $e^{Ht}$ \ \ \ \ \ \ \ \ \ \ \ . \ \ \ \ \ \ \ \ \ \ \ \ \ \ \ \ \ \ \ \ \ \ \ \ \ \ \ \ \ \ \ \ \ \ \ \ \ \ \ \ (11)

\bigskip

$\rho=\rho_{0}$ $e^{-3H\text{ }\left(  1+\beta\right)  \text{ }t}$
\ \ \ \ \ \ \ \ . \ \ \ \ \ \ \ \ \ \ \ \ \ \ \ \ \ \ \ \ \ \ \ \ \ \ \ \ \ \ \ \ \ \ \ \ \ \ (12)

\bigskip

$\phi=\phi_{0}e^{-3H\text{ }\left(  1+\beta\right)  \text{ }t}$
\ \ \ \ \ \ \ \ . \ \ \ \ \ \ \ \ \ \ \ \ \ \ \ \ \ \ \ \ \ \ \ \ \ \ \ \ \ \ \ \ \ \ \ \ \ \ \ (13)

\bigskip

$\beta=-\frac{1}{3}$ \ \ \ \ \ \ \ \ \ \ . \ \ \ \ \ \ \ \ \ \ \ \ \ \ \ \ \ \ \ \ \ \ \ \ \ \ \ \ \ \ \ \ \ \ \ \ \ \ \ \ \ \ \ \ \ \ \ \ \ \ (14)

\bigskip

$p=\beta\rho$ \ \ \ \ \ \ \ \ . \ \ \ \ \ \ \ \ \ \ \ \ \ \ \ \ \ \ \ \ \ \ \ \ \ \ \ \ \ \ \ \ \ \ \ \ \ \ \ \ \ \ \ \ \ \ \ \ \ \ \ \ \ \ (15)

\bigskip

On plugging back the solution (10)-(15), into the field equations, the reader
shall find the conditions among the constants in the problem, namely among
\ \ $\rho_{0}$\ , $\phi_{0}$\ , $\beta$\ , $H$\ , $\omega$ , \ and $\Lambda
$\ . We concentrate our solution into the case \ $k=0$\ \ (a flat Universe),
which is the preference of the inflationary theory (Guth, 1981).

\bigskip

It is evident that we found a similar inflationary solution, for Brans-Dicke
theory, as in the General Relativistic case, for \ $\Lambda\neq0$\ , valid in
the very early Universe.

\bigskip

When \ $\Lambda=0$\ , the reader can check, from the field equations above,
that the following solution applies (notice that we are not imposing, right
now, any equation of state):

\bigskip

$R=D$ $t$ \ \ \ \ \ \ \ \ \ \ \ \ . \ \ \ \ \ \ \ \ \ \ \ \ \ \ \ \ \ \ \ \ \ \ \ \ \ \ \ \ \ \ \ \ \ \ \ \ \ \ \ \ \ \ \ \ \ \ \ (16)

\bigskip

$\phi=C$ $t^{B}$\ \ \ \ \ \ \ \ \ \ \ . \ \ \ \ \ \ \ \ \ \ \ \ \ \ \ \ \ \ \ \ \ \ \ \ \ \ \ \ \ \ \ \ \ \ \ \ \ \ \ \ \ \ \ \ \ \ \ \ (17)

\bigskip

$\rho=\frac{AC}{\alpha D}t^{B-2}$\ \ \ \ \ \ \ \ , and, \ \ \ \ \ \ \ \ \ \ \ \ \ \ \ \ \ \ \ \ \ \ \ \ \ \ \ \ \ \ \ \ \ \ \ \ \ \ \ \ \ (18)

\ \ 

\bigskip$p=St^{B-2}$\ \ \ \ \ \ . \ \ \ \ \ \ \ \ \ \ \ \ \ \ \ \ \ \ \ \ \ \ \ \ \ \ \ \ \ \ \ \ \ \ \ \ \ \ \ \ \ \ \ \ \ \ \ \ \ \ \ \ \ (19)

\bigskip

The constants obey the following conditions, which can be derived from the
field equations,

\bigskip

$S=\frac{AC}{3\alpha}-\frac{1}{24\pi}\left[  B\left(  B+2\right)  \left(
3+2\text{ }\omega\right)  \text{ }C\text{ }\right]  =-\frac{AC}{3\alpha D^{2}%
}\left(  B+1\right)  $\ \ .\ \ \ \ \ \ \ \ \ \ (20)

$1=\frac{8\pi A}{3\alpha D^{2}}+B\left(  \frac{\omega}{6}B-1\right)  $
\ \ \ \ \ \ \ . \ \ \ \ \ \ \ \ \ \ \ \ \ \ \ \ \ \ \ \ \ \ \ \ \ \ \ \ \ \ \ \ \ (21)

\bigskip

$A=\frac{2\omega+3}{2\omega+4}$ \ \ \ \ \ \ \ \ \ \ \ . \ \ \ \ \ \ \ \ \ \ \ \ \ \ \ \ \ \ \ \ \ \ \ \ \ \ \ \ \ \ \ \ \ \ \ \ \ \ \ \ \ \ \ \ \ \ \ (22)

\bigskip

As in the inflationary case, some constants remain undetermined. A negative
pressure, is obviously possible, for a positive energy density, when:

\bigskip

$\frac{A}{3\alpha}<\frac{1}{24\pi}\left[  B\left(  B+2\right)  \left(
3+2\text{ }\omega\right)  \right]  $ \ \ \ \ . \ \ \ \ \ \ \ \ \ \ \ \ \ \ \ \ \ \ \ \ \ \ \ \ \ \ \ \ \ (23)

\bigskip

$\omega>-\frac{3}{2}$ \ \ \ \ \ \ \ \ \ \ \ \ \ \ \ ,

\bigskip

$A>0$ \ \ \ \ \ \ \ \ \ \ \ \ \ \ \ , and,

\bigskip

$C>0$ \ \ \ \ \ \ \ \ \ \ \ \ \ \ \ .

\bigskip

The weak energy condition $\rho>0$\ \ is, thus, obeyed.\ \ \ \ \ 

\bigskip

This model, which entails the case given by the equation of state
($p=-\frac{1}{3}\rho$) , shows a complete accord among Brans-Dicke and General
Relativistic theories; in the former, however, there is more room for
accomodating constants; this is because of the introduction of the
scalar-field $\phi$\ \ , and the existence of a new constant, the coupling one
($\omega$).\ \ \ 

\bigskip

As Berman(2006, 2006a) has remarked for the Einstein's case, the Brans-Dicke
relation (1), has the consequence of driving an exponential inflationary phase
in the very early Universe, not only in General Relativity, but as we have
shown now, also in Brans-Dicke theory.

\bigskip\ \ \ \ \ \ \ \ \ 

{\Large Acknowledgements}

\bigskip

The author gratefuly thanks his intellectual mentors, Fernando de Mello Gomide
and M(urari) M(ohan) Som, and also is grateful for the encouragement by Geni,
Albert, and Paula.

\bigskip

{\Large References}

\bigskip

1. Weinberg, S. - \textit{Gravitation and Cosmology}, Wiley, N.Y., (1972).

\bigskip2. Guth, A. - Phys. Rev. \textbf{D23}, 347, (1981).

3. \bigskip Berman, M. S. (2006) - Chapter 5 of: \textit{Trends in Black Hole
Research}, ed by Paul V. Kreitler, Nova Science, New York.

\bigskip4. Berman, M. S. (2006a) - Chapter 5 of: \textit{New Developments in
Black Hole Research}, ed by Paul V. Kreitler, Nova Science, New York.

\bigskip\bigskip5. Uehara, K.; Kim, C.W. (1982). Physical Review.
\textbf{D26}, 2575.

\bigskip6. Berman, M. S.; Marinho Jr, R.M. (2001) Astrophysics and Space
Science, \textbf{278}, 367.

\bigskip7. Berman,M.S. (1994) - Astrophysics and Space Science \textbf{222}, 235.

\bigskip8. Brans, C.; Dicke, R.H. (1961). Physical Review, \textbf{124}, 925.

\end{document}